\documentclass[twocolumn,trackchanges]{aastex63}
\usepackage{natbib}
\usepackage{hyperref}

\def	\cm		{\,{\rm {cm}}}
\def	\m		{\,{\rm m}}
\def	\km		{\,{\rm {km}}}
\def	\s		{\,{\rm {s}}}
\def	\g		{\,{\rm {g}}}
\def	\kg		{\,{\rm {kg}}}
\def	\Nm		{\,{\rm {N\,m}}}

\shorttitle{interstellar asteroids rotation with mechanical torques}
\shortauthors{W.H.Zhou}

\begin{document}

\title{Oumuamua's Rotation with the Mechanical Torque Produced by Interstellar Medium}

\correspondingauthor{Wen Han Zhou }
\email{u3006384@connect.hku.hk}

\author{Wen Han Zhou}
\affiliation{Department of Physics, The University of Hong Kong\\
Pokfulam Road\\
Hong Kong}

\begin{abstract}

The first interstellar object ‘Oumuamua is discovered in 2017. When 'Oumuamua travels in interstellar space, it keeps colliding with interstellar medium (ISM). Given a sufficiently long interaction time, its rotation state may change significantly because of the angular momentum transfer with interstellar medium.  {Using generated Gaussian random spheres with the {dimension ratios} 6:1:1 and 5:5:1, this paper explores the ISM torque curve and proposes that ISM collision may account for ‘Oumuamua’s tumbling with the simple constant-torque analytical method. The statistic results show that the asymptotic obliquities distribute mostly at $0^\circ$ and $180^\circ$ and most cases spin down at the asymptotic obliquity, indicating the ISM collision effect is similar to the YORP effect with zero heat conductivity assumed. Given a long time of deceleration of the spin rate, an initial major-axis rotation may evolve into tumbling motion under ISM torque. Using a constant-torque analytical model, the timescales of evolving into tumbling for the sample of 200 shapes are found to range from several Gyrs to ens of Gyrs, highly dependent on the chosen shape. The mean value is about $8.5 \pm 0.5$ Gyrs for prolate shapes and $7.3 \pm 0.4$ Gyrs for oblate shapes. Rotation of asteroids in the Oort cloud might be also dominated by the ISM collision effect since the YORP effect is quite weak at such a long distance from the Sun. Although this paper assumes an ideal mirror reflection and a constant relative velocity of 'Oumuamua, the results still show the importance of the ISM collision effect.}

\end{abstract}


\section{Introduction} \label{intro}

The first known interstellar object, 1I/2017 U1 (‘Oumuamua) was discovered by the Pan-STARRS survey in 2017 \citep{M1998}. The light curve of ‘Oumuamua shows that it is extremely elongated with the axial ratio of at least 6:1:1 and it is in an excited rotation around the non-principal axis, which is also referred to as the tumbling \citep{J2017}.  {Recent research shows the oblate shape ($115\times 111\times 19 \m$) also fits the light curve well \citep{M2019}. \citet{B2018} show 'Oumuamua rotate around the shortest axis with a period of $8.67\pm0.34\,\rm{hours}$ and around the long axis with a period of 54.48\,hours. The origin of its tumbling motion is still uncertain.}

Rotational dynamics is an important key to understand the history and evolution of asteroids. For the interstellar asteroid, which is almost isolated in the deep space, even a seemingly unimportant factor can turn out to be dominating on Gyr timescale. In the solar system, several factors, such as (1) Yarkovsky-O’Keefe-Radzievskii-Paddack effect (the YORP effect), (2) the tidal effect caused by nearby massive objects, (3) collisions with other objects and (4) internal energy dissipation due to inelastic deformation, are well known to affect the long-term dynamical evolution of asteroids. But they are ineffective in the case of the interstellar asteroid. Appendix \ref{A} discusses the major limits of these factors in the case of interstellar asteroids and explains why this paper ignores them. Besides these mentioned factors, interstellar medium, although extremely thin in deep space, could change the rotation state significantly, given a sufficiently long interaction timescale. The mechanical torque produced by interstellar gas flow has been proved to change the rotation of interstellar grains. \citet{LH2007} state that for an irregular interstellar grain, the external mechanical torques can spin up the grain and lead to a suprathermal rotation. According to \citet{HTLA2019}, large dust grains can be spinned up to disruption when the centrifugal stress due to rotation is larger than the maximum tensile stress that the dust can hold. As for interstellar asteroids, \citet{HLLC2018} propose that the interstellar asteroid may spin up and get disrupted under the mechanical torques produced by interstellar gas.

Can the almost perfect vacuum in interstellar environment yield a dominating effect on the rotation of the asteroid on a meaningful timescale? The answer is still unclear.  Using random walk formula, \citet{HLLC2018} give a rough estimate of the asteroid’s lifetime, which is relevant to the size and a shape parameter $N_{\rm fc}$, before the asteroid gets disrupted under the ISM torque. However, since the typical value of shape parameter $N_{\rm fc}$ (measuring the shape asymmetry of the body) is unclear, the timescale of this the ISM collision effect is not fully answered. This paper introduces the Gaussian random sphere model and the torque curve, which were successfully used to explain the mechanism of the YORP effect \citep{VC2002}, to the topic of the interstellar asteroid. To obtain a statistic description, 100 random prolate shapes and 100 oblate shapes are generated in this paper. The results, as shown in Section \ref{stat}, indicate the asteroids are much more likely to spin down at asymptotic obliquities, {which partially question the assumption of spinning up to disruption, together with the estimate of the asteroid’s lifetime in \citet{HLLC2018}.} In the case of spinning down, the asteroid may finally turn to tumbling, which provides an explanation of the tumbling state of ‘Oumuamua. To give an estimate of the timescale for evolving tumbling, a simple constant-torque is applied, and the statistic result of 200 shapes show this process is on Gyr timescale. Besides, the ISM collision effect is expected to be similar to the YORP effect in some regards, both of which are based on the torque on the surface. Thus this paper also explores the similarity and difference between the ISM collision effect and the YORP effect on the torque curve and statistic behaviors.

 {To give an approximation for the ISM collision effect, a simple full mirror reflection is assumed. During a small time interval $\Delta t$, the mass of hydrogen atoms colliding with the small area $\Delta S$ is }
\begin{equation}
\Delta m_\mathrm{p} = m_\mathrm{p}n_\mathrm{p}\Delta S v \Delta t, \label{E1}
\end{equation}
 where $v$ is the speed of the asteroid relative to interstellar gas, $m_\mathrm{p}$ is the mass of proton, and $n_\mathrm{p}$ is the number density of protons. According to \citet{KD2008}, the average number density of hydrogen atoms at the solar radius in the mid-plane galaxy is around $0.9 \cm^{-3}$. In this paper, I take a reasonable value $1\cm^{-3}$ to be the averaged number density of protons. {Collisions with dust particles are neglected} here since the typical gas-to-dust mass ratio in the galaxy is about 100: 1 \citep{BL2018}. After a frontal impact, the angular momentum the particles transfer to the surface area is 
\begin{equation}
\Delta L = 2 \Delta m_\mathrm{p} vl, \label{E2}
\end{equation} 
 with $l$ symbolizing the distance from the surface area to the asteroid’s barycenter. Substituting Equation (\ref{E1}) into Equation (\ref{E2}), the torque exerted on this small area is obtained by
\begin{equation}
T = {\Delta L \over \Delta t}= 2m_\mathrm{p}n_\mathrm{p}\Delta S  v^2l. \label{E3}
\end{equation}
Summing up the torques over the whole surface, a non-zero net torque will be produced if the shape of the interstellar asteroid is asymmetric. The torque component along the major principal axis of the body can change the spin rate of the body. Assuming the asteroid has a mean radius of $R$, the net torque along the major axis can be obtained by substituting $\Delta S \sim R^2$ and $l \sim R$ into Equation (\ref{E3}) and multiplying Equation (\ref{E3}) by a coefficient $C_\mathrm{z}$ which measures the asymmetry:
\begin{equation}
T_z = 2m_\mathrm{p}n_\mathrm{p}v^2R^3C_\mathrm{z},\label{T_z}
\end{equation}
 {a similar form to Equation (3) in \citet{GS2019}. \citet{GS2019} shows that for the YORP torque, $C_\mathrm{z}$ is around 0.01 to Type $\rm{\uppercase\expandafter{\romannumeral1}}$/$\rm{\uppercase\expandafter{\romannumeral2}}$ asteroids. Note that the ISM collision effect is not the same thing as the YORP effect, but for an order-of-magnitude estimate of the torque, an analogy can be made. Thus here I estimate $C_\mathrm{z}$ to be 0.01, which will be revisited in Section \ref{stat}. Plugging $m_\mathrm{p} = 1.67 \times 10^{-27}\kg$, $n_\mathrm{p}=1 \cm^{-3}$, $v = 20\km\,\rm{s}^{-1}$, $R=60\m$ into the Equation (\ref{T_z}), we have $T_z = 2.89\times 10^{-9} \Nm$. Further, the formula describing the uniform rotation about the major axis \citep{R2000} is given by}
\begin{equation}
I_z{d\omega \over dt} = T_z,
\end{equation}
 {where $I_z$ is the maximum moment of inertia of the asteroid. Assumed as $I_z  =8\pi\rho R^5/15$ for a sphere with $\rho =1500\kg\,\rm{m}^{-3}$, we can obtain the angular acceleration $d\omega/dt = 1.5\times10^{-21} \,\rm{rad}\,\rm{s^{-2}}$. Therefore for a rotator with a period of 8 hours, the angular velocity will double in about 4 Gyrs given a positive torque. It is reasonable to speculate that this effect will dominate the rotation on the timescale of Gyrs for the interstellar asteroid with a mean radius of tens of meters. }

The mechanism of the ISM collision effect is similar to that of the YORP effect to some extent, both of which are based on the force normal to the surface. { {As it will be shown later in the article}, the ISM collision effect is more like the YORP model with zero thermal conductivity.} the YORP effect is believed to influence the rotation evolution of asteroids such as changing the obliquity and angular velocity \citep{R2000,VC2002,S2007}, and even causing the tumbling motion \citep{VBNB2007}. {Combined with the effect of internal energy dissipation, the YORP effect may lead to a new asymptotic tumbling state of asteroids \citep{BM2015}. }

In Section \ref{simulation}, the main equations and the shape model in the simulation are introduced. Highly elongated and disk-like irregular shapes are generated as pseudo ‘Oumuamua. In Section \ref{result}, the description of the torque curve and statistic results are shown. The origin of tumbling motion of the interstellar asteroid ‘Oumuamua is discussed. A comparison is made between the ISM collision effect and the YORP effect for distant asteroids in the Solar system.

\section{Simulation}
\label{simulation}
\subsection{Equations} 
In this study, the kinetic approach is used to analyze the collisions between the interstellar asteroid and gas, {as first proposed by \citet{HLLC2018}.}

 {For a surface element dS, The mass of particles colliding with the surface within a time interval dt is}
\begin{equation}
dm_\mathrm{p} = m_\mathrm{p}n_\mathrm{p}({\bf {v \cdot  n} } )dS dt. \label{dMp}
\end{equation}
 Here I set $\bf v$ as the translational velocity vector of the asteroid relative to surrounding interstellar gas. Although the asteroid is rotating, the rotational velocity is too small compared to the translational velocity. {Here}, $\bf n$ is the outward unit normal vector of the surface element $dS$. Several assumptions are made in the following derivation: (1) mirror reflection of particles and no sputtering on the surface, (2) the strongly supersonic regime of the body, and (3) the convexity of the surface. After colliding with the surface element, gas particles are reflected, with the velocity component perpendicular to the surface reversed. In this process, the momentum transferred from the particles to the asteroid is 

\begin{equation}
{\bf dp} \approx -2  ({\bf v \cdot n}) {\bf n} \gamma_r dm_\mathrm{p}. \label{dp}
\end{equation}
$\gamma_r$ is the reflection coefficient, with $\gamma_r = 1$ representing a mirror reflection. Plugging (\ref{dMp}) into (\ref{dp}), we can obtain:
\begin{equation}
{\bf df} = {{\bf dp}\over dt} = -2m_\mathrm{p}n_\mathrm{p}({\bf v \cdot n)(v \cdot n)n} \gamma_r dS .    \label{df}
\end{equation}

In this paper $\gamma_r$ is set to be 1 because of the assumption (1). Note that the process of {``sputtering''} also produces a torque, which is calculated to be smaller by one order of magnitude than the ISM collision torque, which is shown in Appendix \ref{B}. We should take caution in further research when neglecting this ``sputtering" effect since the variation by just one order of magnitude could be removed by the uncertainty of the parameters. Assumption (2) discards the effect of the gas particles' thermal motion. Let's consider interstellar gas in the warm phase with the temperature of 5000\,K. Applying Maxwell's velocity distribution, the averaged speed of the gas is $\bar v_\mathrm{gas} = (8k_{\rm B}T/\pi m )^{1/2} \approx 9\km/\rm{s}$, less than the assumed speed of the asteroid. The asteroid can be considered to be supersonic. In the hot region (e.g. above 20000K), the above equations are certainly invalid. Assumption (3) avoids the shadowing and secondary collisions of particles. But we could expect Equation (\ref{df}) is still valid for moderately concave shapes. According to \citet{GKKS2016}, for the YORP effect, the formulas proposed under the assumption of convexity still works well in the case of moderately concave shapes. Considering that the ISM collision effect is similar to the YORP effect in some regards, this rule is expected to be also applicable to the ISM collision effect.

 {The total torque exerted on the whole surface of the body is}
\begin{equation}
{\bf T} = \int \bf r \times df.\label{T}
\end{equation}
{As can be seen from Equation (\ref{df}) and (\ref{T}), if the body is spherical, $\bf r$ is parallel to $\bf n$, resulting in that the term $\bf r \times {df}$ vanishes all over the surface. In this case there will be no torque on the body. In some other special cases, depending on the symmetry of shape and the obliquity, torques exist on the surface but they cancel each other, resulting in zero net torque on the body. However, generically the torques do not cancel each other for an irregular shape and therefore change the body’s rotation.}

\subsection{Shape Model} 
Gaussian random sphere model is {introduced by \citet{M1998}} and has been proved useful in simulating dynamics of asteroids and comets with irregular surface \citep{VC2002}. Considering a spherical coordinates system with the origin at the mass center of the body, the distance from the origin to the point on the surface is expressed as a function of the azimuthal angle $\phi$ and polar angle $\theta$,
\begin{equation}
r_{s}(\theta,\phi) = a (1+\sigma^2)^{-1/2}e^{\omega(\theta,\phi)},\label{r_s}
\end{equation}
where $a$ is the characteristic dimensional factor, $\sigma$ is the variance, measuring the amplitude of deviation from sphericity. In the Equation (\ref{r_s}),
\begin{equation}
\omega (\theta,\phi) = \sum_{l=1}^{l_{max}} \sum_{m=0}^{l}P_l^m(\cos\theta)[a_{lm}\cos(m\phi)+b_{lm}\sin(m\phi)],
\end{equation}
where $P^m_l$  is the Legendre functions, the coefficients $a_{lm}$ and $b_{lm}$ are independent Gaussian random variables with zero mean and variance $\beta _{lm}^2$ given by
\begin{equation}
\beta _{lm}^2 = (2-\delta_{m0}){(l-m)! \over (l+m)!}c_l ln(1+ \sigma ^2).
\end{equation}
Here $\delta_{m0}$ is the Kronecker symbol. {Parameter $c_l$} is given by
\begin{equation}
{c_l = l^{- \alpha} \left(\sum_{l=1}^{l_{\rm{max}}}l^{-\alpha}\right)^{-1},}
\end{equation}
where $\alpha$ tells the angular scale of the deviations.

\begin{figure}[h]
\centerline{\includegraphics[width=\columnwidth]{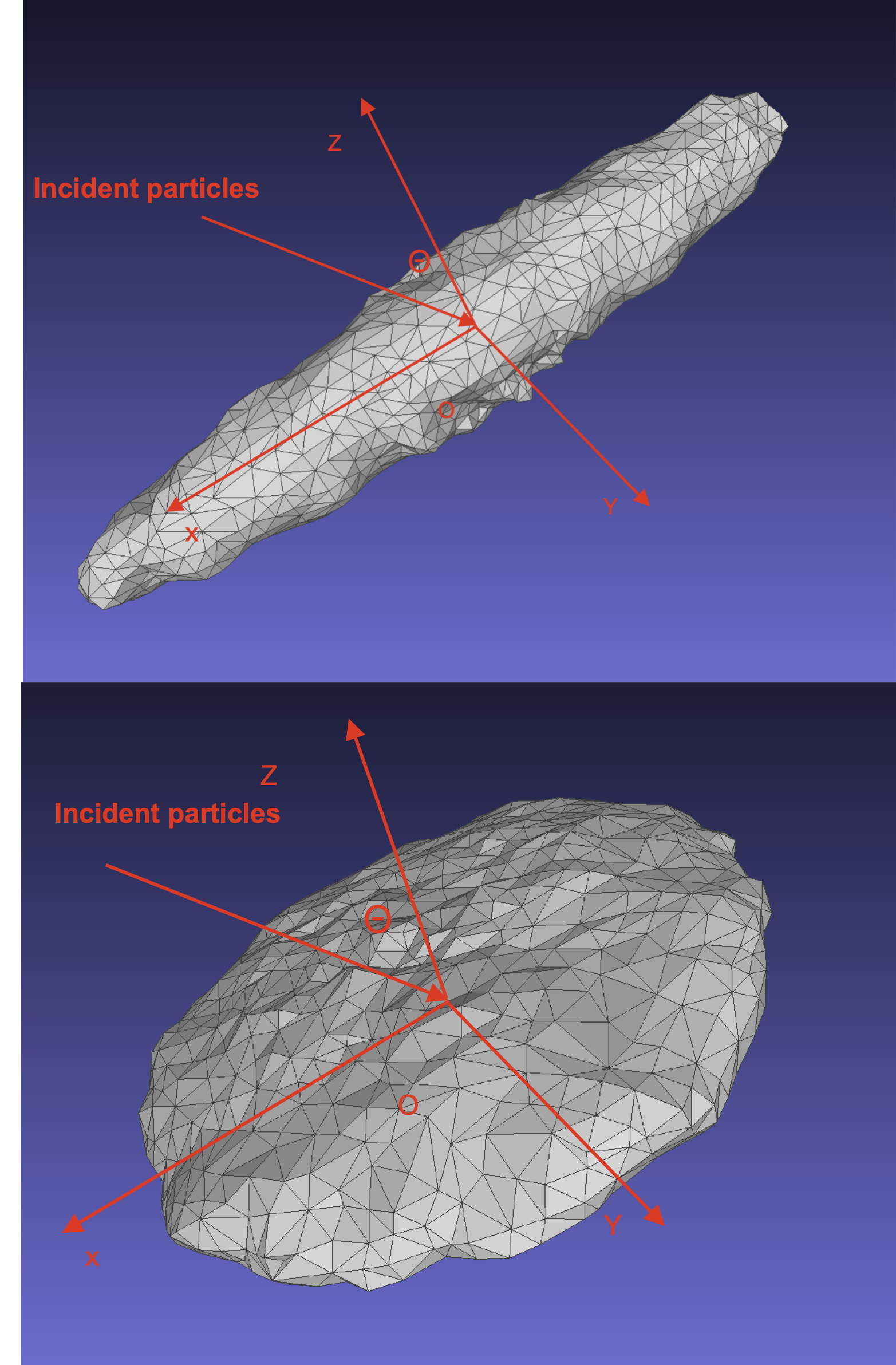}}
\caption{ Two examples of generated random shapes. {The upper shape is a prolate shape with $C/B/A$ = 5.90/1.03/1 and the lower one is an oblate shape with $C/B/A$ = 4.95/4.90/1 where $A,B,C$ are the semi-major axes of the homogeneous ellipsoid with the same moments of inertia, and $A<B<C$.}}
\label{shape}
\end{figure}

In this study I set $l_{\rm max}$ equal to 10,  which is sufficiently large for simulation \citep{DW2016}. I generate 100 highly elongated shapes and 100 disk-like shapes as the examples of pseudo ‘Oumuamua, using about one thousand facets for each object. The dimension ratios of prolate shapes are set about 6:1:1,  together with a mean radius of 60\m, suggested by \citet{J2017} while the oblate shapes have an axis ratio of about 5:5:1 with a mean radius of 31\m, which shows a good fit to the light curve of 'Oumuamua \citep{M2019}. Figure \ref{shape} shows two examples of pseudo ‘Oumuamua. A basic assumption is that the shape of the interstellar asteroid doesn't change noticeably. Although the interstellar medium causes the accretion and erosion on the surface of the object, calculation in Appendix \ref{B} shows the radius change rate is $dR/dt \approx -0.0014\m\,\rm{Gyr}^{-1} $ and can be ignored.

{\section{{Results and Discussions}}\label{result}}
 {\subsection{Torque Curve and Rotation Evolution}}
{The averaged torque produced by ISM collision over one spin period is an important quantity when we look into the long-term evolution of the interstellar asteroid. In the following text,} the coordinate system is body-fixed such that the object has the moments of inertia of $(I_x,I_y,I_z)$ where $I_x<I_y<I_z$. The obliquity $\varepsilon$ is defined as the angle between $\bf e_z$ direction and the object’s velocity with respect to the interstellar medium, $ \cos\varepsilon = \bf e_z \cdot \bf e_v$ with $\bf e_v = \bf v/|\bf v|$. In this paper I prefer to define 
\begin{equation}
T_z = \bf T \cdot \bf e_z,
\end{equation}
\begin{equation}
T_\mathrm{\varepsilon} = \bf T \cdot \bf {e_{\bot}},
\end{equation}
{as indicators of the effect, where ${\bf e_{\bot} }= ({\bf e_z} \cos\varepsilon - {\bf e_v})/\sin\varepsilon$, slightly different from the variables used by \citet{VC2002}. $T_z$ and $T_\mathrm{\varepsilon}$ strictly rely on the orientation and velocity of the object, which means that when spin direction reverses, the $T_z$ and $T_\mathrm{\varepsilon}$ components do not change their signs, but result in opposite effects (e.g. changing from acceleration to deceleration or inclining to the inverse direction). The torque curve reveals the dependence of $T_z$ and $T_\mathrm{\varepsilon}$ on the obliquity $\varepsilon$ in the range of $(0^\circ,180^\circ)$.}

{Unlike in the case of the YORP curve, the torque curve here has no properties of symmetry or  anti-symmetry for $T_z$($\varepsilon$) and $T_\mathrm{\varepsilon}(\varepsilon)$. This is because the ISM torque is only averaged over the spin period, while in the case of the YORP effect the periodical rotation around the Sun leads to periodical illumination, which gives the average torque properties of symmetry and antisymmetry. }

\begin{figure}[h]
\centerline{\includegraphics[width=\columnwidth]{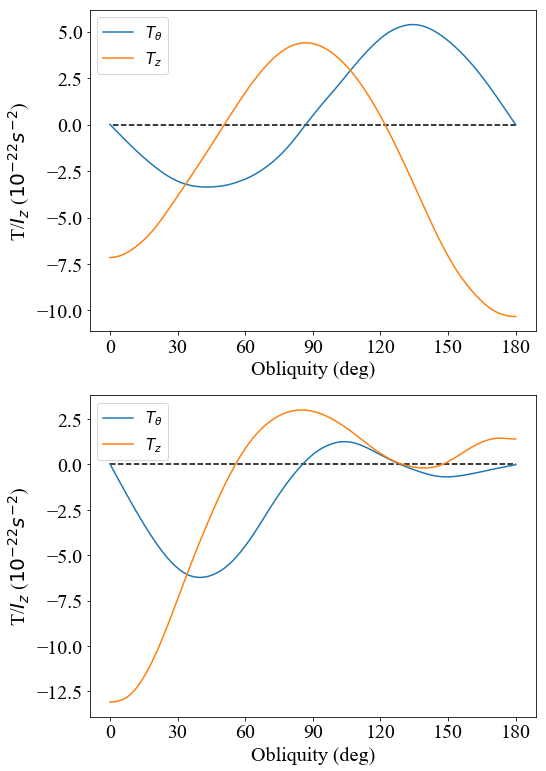}}
\caption{ The ISM torque curves of two generated shapes. The blue line denotes the torque component $T_\mathrm{\varepsilon}$ and the orange line denotes $T_z$. The black dash is auxiliary line located in $T = 0$.}
\label{torque}
\end{figure}

{Figure \ref{torque} shows different torque curves. When the obliquity is $0^\circ$ or $180^\circ$, $T_\mathrm{\varepsilon}$ vanishes due to the procedure of average over one spin period while $T_z$ does not. In other values of obliquity $\varepsilon$, the non-zero torque component $T_\mathrm{\varepsilon}$ leads to the inclination of the body. The direction of inclination is dependent on the sign of $T_\mathrm{\varepsilon}$ as follows: A negative value of $T_\mathrm{\varepsilon}$ will decrease the obliquity $\varepsilon$, moving the spin axis towards the velocity vector while a positive one will increase the obliquity $\varepsilon$.  Thus a zero point of $T_\mathrm{\varepsilon}$ may indicate a stable orientation of the asteroid, where the body may maintain a slow precession of the spin axis around the velocity vector. On the other hand, the $T_z$ tells the spin acceleration. }

{The torque curve here is slightly different from the YORP curve because of the difference brought by asymmetry. The torque curve has to be considered in the obliquity range $[0^\circ, 180^\circ]$, and $T_\mathrm{\varepsilon}$ is not always equal to 0 at $\varepsilon$ = $90^\circ$. Previous research on the YORP curve divides the majority of asteroids into four types according to the torque component $T_\mathrm{\varepsilon}$ \citep{VC2002}. With the assumption of principal-axis rotation, asteroids of Type $\rm{\uppercase\expandafter{\romannumeral1}}$ will reach a stable orientation at $\varepsilon$ = $90^\circ$ ,and asteroids of Type $\rm{\uppercase\expandafter{\romannumeral2}}$ will asymptotically reach $\varepsilon$ = $0^\circ$ or $180^\circ$. Asteroids of Type $\rm{\uppercase\expandafter{\romannumeral3}}$ and Type $\rm{\uppercase\expandafter{\romannumeral4}}$ both have $T_\mathrm{\varepsilon}$ = 0 at some values of $\varepsilon$ in range ($0^\circ$, $90^\circ$) or ($90^\circ$, $180^\circ$), but Type $\rm{\uppercase\expandafter{\romannumeral3}}$ asteroids will get stable at the zero point in the ($0^\circ$, $90^\circ$) or ($90^\circ$, $180^\circ$) obliquity range while Type $\rm{\uppercase\expandafter{\romannumeral4}}$ asteroids’ stable state has multiple answers dependent on the initial obliquity. The final stable obliquity is also called asymptotic obliquity \citep{VC2002}. Thus there is no need to introduce another classification system for the small difference stated above. Readers can easily find the asymptotic obliquities in the torque curves by looking for the nodes where $T_\mathrm{\varepsilon}(\varepsilon)$ = 0 and $T'_\varepsilon(\varepsilon)<0$. }

{For example, the upper curve in Figure \ref{torque} has three zero points of $T_\mathrm{\varepsilon}(\varepsilon)$. Let’s call them $\varepsilon_1$, $\varepsilon_2$, $\varepsilon_3$. As discussed, $\varepsilon_1 = 0^\circ $ and  $\varepsilon_3 = 180^\circ $. As we can see, $T_\mathrm{\varepsilon}(\varepsilon_2)= 0$, $T'_\varepsilon(\varepsilon_2)>0$. Therefore, only obliquity $\varepsilon_1$ and obliquity $\varepsilon_3$ are possible final stable points, dependent on the initial obliquity. If the initial obliquity is in range of ($\varepsilon_1$, $\varepsilon_1$), the asteroid will reach $\varepsilon_1$ ultimately, and if the initial obliquity is in range of ($\varepsilon_2$,$\varepsilon_3$), the asteroid will reach $\varepsilon_3$. What’s more, the $T_z(\varepsilon_1)$ and $T_z(\varepsilon_3)$ are both negative, meaning that the asteroid will keep decelerating at these stable obliquities. The lower example has two asymptotic obliquities, $0^\circ$ and $126^\circ$. Readers may refer to \citet{S2007} for an analytical method on the effect of the torques.}

 {\subsection{Statistical Results} \label{stat}}
{Considering that the ISM torque is highly dependent on the specific shape of the asteroid, 200 Gaussian random shapes are generated to obtain a statistical description. According to two different opinions about 'Oumuamua's shape, half of the sample are set to have a dimension ratio of about 6:1:1 with an equivalent radius of about 60\m \citep{J2017} and the other half have a dimension ratio of around 5:5:1 with an equivalent radius of 31\m \citep{M2019}.

\begin{figure}[h]
\centerline{\includegraphics[width=\columnwidth]{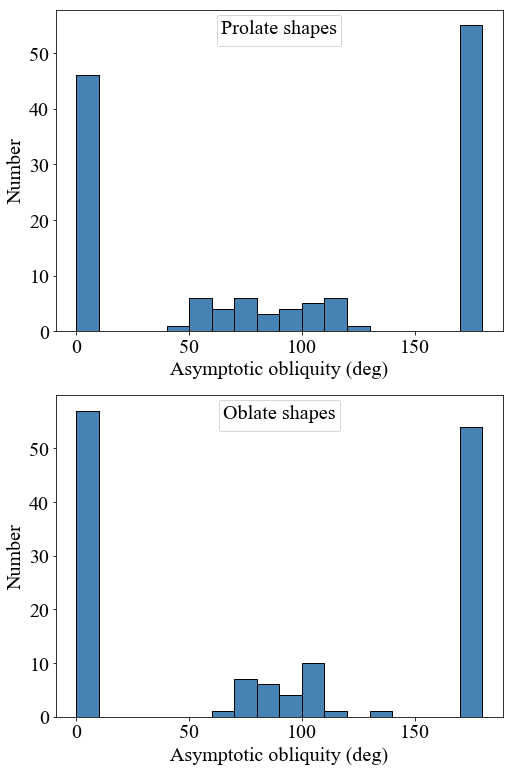}}
\caption{ Distributions of the asymptotic obliquities of ISM torques for the sample of 200 shapes. The upper figure is the result for prolate shapes and the lower one is for oblate shapes}\label{obliquity}
\end{figure}

The distribution of asymptotic obliquities is shown in Figure \ref{obliquity}. There are 137 asymptotic obliquities in total for 100 tested prolate shapes and 141 for oblate shapes. As we can see, the majority of the asymptotic obliquities are located at $0^\circ$ and $180^\circ$ with nearly equal likelihood (Note that obliquity $0^\circ$ and $180^\circ$ are separate). {At the current stage, $C_z$ in Equation (\ref{T_z}) is easily calculated by substituting asymptotic obliquities into Equation (\ref{T_z}). For prolate shapes, $C_z =0.036\pm0.002$ and for oblate shapes, $C_z = 0.0057 \pm 0.0004$, which indicates prolate shapes are more sensitive to the ISM torques.}

{Another important statistic result is that the cases of deceleration at the asymptotic obliquity in the tested shapes are {much more numerous than} the cases of acceleration. Among the 200 shapes, only five show acceleration of rotation at the asymptotic obliquity, which means the interstellar asteroids are much more likely to asymptotically decelerate their spin rates under ISM torque. Similar statistical results occur in the research on the YORP torque with the assumption of zero thermal conductivity. \citet{CV2004} {show} that, in the limit of zero thermal conductivity, most cases {decelerate} at the asymptotic obliquity, while bodies with finite thermal conductivity equally accelerate and decelerate their spin rates. It is seen that the ISM collision effect is more similar to the YORP effect with the zero heat conductivity assumed.}

\subsection{Despin and Tumbling}\label{despintumbling}
{A fundamental question about ‘Oumuamua is how the tumbling motion arises and the answer is still unclear. Usually, possible reasons for the tumbling motion of asteroids include (1) original tumbling when the asteroid gets separated from its parent body, (2) Impacts in the home system \citep{HP2013}, (3) the YORP torques\citep{VBNB2007}, (4) tidal torques \citep{S2001,SJB2006}. In the case of an interstellar asteroid, another mechanism might account for the tumbling: collision with interstellar medium during its long-time travel in interstellar space.}

{Most cases in the simulation keep decelerating the rotation rates after they reach the asymptotic obliquity. When the spin rate is low enough, even small external torque like ISM collision torque could trigger the tumbling motion. In this study, this possible explanation for the tumbling motion of ‘Oumuamua is not examined directly through simulation due to the extremely large timescale of this effect and the limited computer power. However, considering the similarities of the “torque” mechanism and statistical behavior between the ISM collision effect and the YORP effect (zero heat conductivity assumed), we can follow the work on the YORP effect on the tumbling motion of asteroids in the Solar system \citep{VBNB2007,BRV2011}. According to \citet{VBNB2007}, provided that the rotation is long enough, the YORP torque that initially decelerates the spin rate could deviate the body from the initial rotation about the principal axis and turn the body into an asymptotic state of tumbling.}

\begin{figure}[h]
\centerline{\includegraphics[width=\columnwidth]{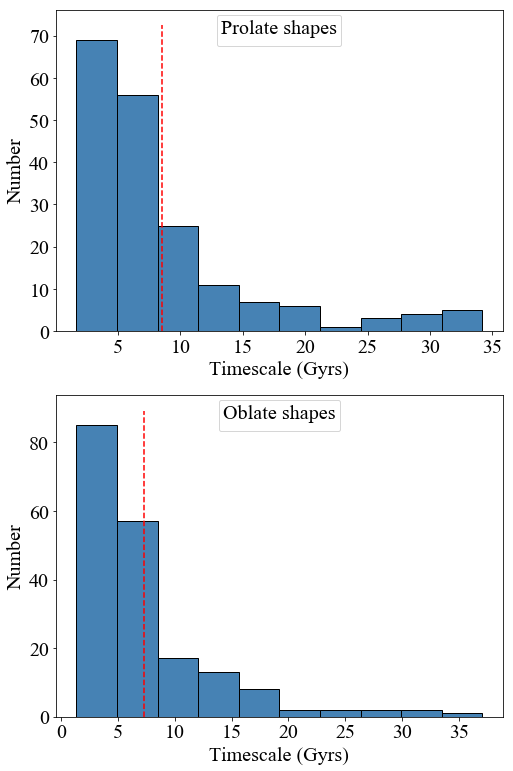}}
\caption{ The distribution of the timescale of evolving into tumbling from SAM mode for 200 pseudo ‘Oumuamua models with initial spin period of 8.67 hours. The upper figure is the result for prolate shapes and the lower one is for oblate shapes. The mean values, denoted by the red dash lines, are $8.5 \pm 0.5$ Gyrs and $7.3 \pm 0.4$ Gyrs, respectively. }
\label{tumbling}
\end{figure}

{While it is difficult to completely describe how external torques excite the tumbling in an analytical way, a theory assuming a constant torque \citep{TL1991} is found to be in good agreement with simulation results of the YORP torques, according to \citet{VBNB2007}. Here we use this method to analyze the effect of ISM torques. The point here is to identify a constant torque component ${\bf T} = (T_x,T_y,T_z) \ne \bf 0$. Suppose that the interstellar asteroid initially rotates around the axis of maximum moment of inertia (SAM mode), and have an initial stable obliquity around {$0^\circ$ or $180^\circ$}. At this special obliquity, the averaged torque in the body-fixed frame does not change when the body rotates. Therefore, the constant-torque model can be used. In a first-order approximation, {the $z$ component of the angular momentum vector ${\bf L} = (L_x,L_y,L_z)$ is expressed as: }}
\begin{equation}
L_z(t) = L_z(0) + T_zt.
\end{equation}
{The initial conditions are $L_x(0) = 0, L_y(0) = 0, L_z(0)>0$. If $T_z > 0$, the $L_z(t)$ increases unlimitedly and $L_x(t)$ and $L_y(t)$ tend to oscillate with some constant amplitudes, leading to a stable SAM mode rotation, until it {gets} disrupted. If $T_z<0$, the $L_z(t)$ decreases while the amplitudes of oscillation of $L_x(t)$ and $L_y(t)$ increases, finally evolving into tumbling state. \citet{VBNB2007} introduce a formula to estimate a typical timescale for a SAM mode body to enter tumbling state:}
\begin{equation}
\Delta t = - {L_z(0) \over T_z}.\label{t}
\end{equation}

{Using Equation (\ref{t}), the evolving tumbling timescale of 200 generated shapes are calculated with an initial spin period of 8.67 hours assumed. Figure \ref{tumbling} shows the distribution of the timescale required for an initial SAM mode rotation to evolve into tumbling. {The mean timescale is about $8.5 \pm 0.5$ Gyrs for prolate shapes and $7.3 \pm 0.4$ Gyrs for oblate shapes, which is of the same order of magnitude as our estimation in Section \ref{intro}.} It should be noted that the timescale is highly dependent on the chosen shape and varies a lot among different shapes, as indicated in Figure \ref{tumbling}. The mean timescale should be used carefully as currently we do not know the exact shape of ‘Oumuamua due to the limited observation data. We expect the ISM effect to be sensitive to the {surface topography}, in view of extreme sensitivity of the YORP effect to fine structure on the surface of asteroids \citep{S2009}.} As the angular acceleration scales as $\sim 1/R^2$, the effect of the torque is more intensive for smaller bodies.

\subsection{Implications for asteroids in the Oort cloud}
\label{implication}

The way ISM affects asteroids’ rotation is quite similar to the YORP effect, both of which are based on torques on the surface. A comparison of the magnitude between these two torques may help know more about asteroids in the Solar system. To remove the influence of the shape, here I compare the ratio of the force to the surface element $df/dS$. According to \citet{VC2002}, the recoil force due to the thermally emitted radiation from the surface element is given by
\begin{equation}
{\bf df} \sim -{ 2 \epsilon \sigma T^4 \over 3c_0} {\bf n} dS \sim -{2\Phi \over 3c }({\bf n \cdot n_0}) {\bf n} dS. \label{dfYORP}
\end{equation}
{Here}, $\bf n$ is the outward normal vector to the surface element. $\bf n_0$ is the direction from the object to the Sun.  And $\Phi$ is the solar flux at the distance of the object from the Sun. { To remove the influence of obliquity, I assume $\bf n \cdot \bf {n_0} = 1$ in Equation (\ref{dfYORP}), $\bf n \cdot \bf e_v = 1$ in Equation (\ref{df}). Thus the Equation (\ref{dfYORP}) and Equation (\ref{df}) are rearranged as:
 \begin{equation}
 {\left (df \over dS\right )}_{\rm YORP} ={2\Phi \over 3c} = {2\Phi_0 \over 3c}\left({d_\oplus \over d_{\rm{A}}}\right)^2,
 \end{equation}
  \begin{equation}
 {\left (df \over dS\right )}_{\rm ISM} = 2m_{\rm p}n_{\rm p}v_{\rm{Oort}}^2 \label{E20}.
 \end{equation}
Here, $d_\oplus$ and $d_{\rm{A}}$ are the distances to the Sun from the Earth and the asteroid, respectively. $\Phi_0 = 1367\,\rm W\, \rm m^{-2}$ is the solar constant. $v_{\rm{Oort}}$ in the Equation (\ref{E20}) denotes the relative speed of the asteroid in the Oort cloud. \citet{S1990} suggests adopting $v = 20 \km\,s^{-1}$ (the typical solar velocity relative to local ISM ) for the velocity of asteroids in the Oort cloud. The orbital velocities are neglected since they are much smaller and almost averaged out over one period.}

 \begin{figure}[h]
\centerline{\includegraphics[width=\columnwidth]{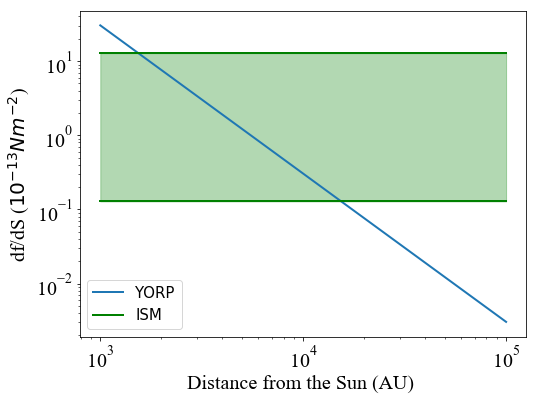}}
\caption{  {Comparison between ISM torque (green region) and the YORP torque (blue line).  The ISM torque is estimated under the number density of interstellar gas ranging from 0.01$\,\mathrm{cm^{-3}}$ to 1$\,\mathrm{cm^{-3}}$. The orbital speed relative to interstellar gas is assumed to be 20$\km\,\rm s^{-1}$}}
\label{comparison}
\end{figure}

 As Figure \ref{comparison} shows, the ISM collision effect is much weaker than the YORP effects when discussing objects in the {main asteroid belt}. However, since the YORP effect declines very quickly with distance to the Sun increasing, the ISM collision effect may dominate extremely distant asteroids’ rotation, such as asteroids in the Oort cloud, which can be as far as about {100000\,AU} away from the Sun. It is suggested that one to two percent of the Oort cloud population are asteroids \citep{WL1997}.  {Further research on the ISM collision effect might give some constraints on their rotation states, {depending on their shapes and sizes.}}

 \section{Conclusions}
When interstellar asteroids travel in interstellar space, they keep colliding with interstellar medium. Although the effect is very weak, when considering a large timescale as billions of years, it can change the rotation of the asteroids considerably.  {In this paper, the torque curve of the ISM collision effect is studied and a mechanism accounting for ‘Oumuamua’s tumbling is proposed. The main findings and discussions include:}

 {(1) The body is much more likely to spin down at the asymptotic obliquities that are mainly distributed at $0^\circ$ and $180^\circ$ with nearly equal likelihood, according to the statistic result. In this dimension, ISM torque is more like the YORP torque with zero thermal conductivity model than that with a finite-conductivity model. }

 {(2) ‘Oumuamua may be born with a major axis rotation and turn into tumbling under the effect of the ISM torque. The timescale of this process, calculated from 200 pseudo ‘Oumuamua generated by Gaussian random sphere method, ranges from several Gyrs to tens of Gyrs, depending on the chosen shape. The mean value is $8.5 \pm 0.5$ Gyrs for prolate shapes and $7.3 \pm 0.4$ Gyrs for oblate shapes.}

 {(3) the ISM collision effect might also dominate rotation of distant small asteroids in the Oort cloud as the dim light extremely weakens the YORP effect.}

\acknowledgements
I would like to thank my supervisor Meng Su, and Xiaojia Zhang for discussions during the early stage of this study, Michael Efroimsky for discussions regarding the asteroid's internal energy dissipation, Xiaoran Yan for discussions about polyhedron modeling and the YORP effect, Yun Zhang for help with disruption of rubble piles. Finally, I am deeply thankful to the reviewer, who gives a lot of professional and valuable suggestions with a comprehensive and detailed report.

\appendix
\section{Other effects of interstellar environment}
\label{A}

{In the interstellar environment, the rotation of the interstellar asteroid might also be affected by some factors other than interstellar medium. }

Collisions between asteroids may dominate the rotation evolution of asteroids in the Solar system, but it becomes doubtable in the case of interstellar asteroids. The number density of interstellar objects like ‘Oumuamua or larger is estimated to be about $0.1\,\rm{AU}^{-3}$ \citep{DTT2018}. The mean free path of a collision between two ‘Oumuamua-like objects is $l \sim 1 / \pi R^2 n_{\rm{ISO}} $. Considering the velocity of ‘Oumuamua is $20\,\km\s^{-1}$, the expected mean collision time is $\tau_\mathrm{c} \sim {l / 2v} \sim 10^{14}$ years, much longer than the timescale considered in this paper.  {However, there is no enough data to estimate the number density of objects smaller than ‘Oumuamua by an order of magnitude or more, which could also cause a big change of the rotation rate after impacting ‘Oumuamua. In the current stage, without enough background information, this paper ignored the collision between interstellar objects. }

The internal energy dissipation will cause a tumbling asteroid to be in alignment to the axis of maximum moment of inertia \citep{BSG1973,LE1999}. \citet{F2018} estimate the damping time of ‘Oumuamua longer than $4 \times 10^{10}$ years, using $Q$ model with $Q$ in range of (100, 1000). \citet{D2018} state that the damping timescale of ‘Oumuamua is estimated to be at least 1 Gyr.  {As for a tumbling object, the mechanism of internal energy dissipation and external torques like the YORP effect may cause an asymptotic state of stationary tumbling with a fixed rotation period \citep{BM2015}. If ISM collision and strong internal energy dissipation are assumed, the interstellar asteroid may evolve into some special tumbling state under both effects. However, the effect of internal energy dissipation may be dramatically weak in deep space. {\citet{FE2017} found the damping timescale $\tau_{\rm damp} \varpropto \eta$ where $\eta$ is the viscosity of the body. Roughly, $\eta \varpropto e^{1/T}$ if the Arrhenius's law is applied. Thus, the low temperature of the interstellar asteroid may lead to the extremely large damping timescale.} {\citet{K2020} states the damping timescale for 'Oumuamua can be surprisingly as large as $10^{23}~10^{193}$ years.} If the typical timescale of ISM collision is shorter than damping timescale, the effect of internal energy dissipation can be ignored, as Section \ref{despintumbling} does when estimating the timescale of evolving tumbling. On the other hand, the problem brought by the weak internal energy dissipation is the invalidity of the assumption of principal axis rotation in the analysis of long-term rotation evolution. This paper does not discuss the internal energy dissipation of 'Oumuamua in detail.}

 {Gravitational torques due to the galactic center may change the rotation of interstellar asteroids. Previous work shows that in the Solar system, the gravitational torque due to the Sun affects the orientation of the asteroid in the long term \citep{VC2002}. Assuming a fixed orbit, the torque causes a spin axis precession around the normal to the orbit plane on the timescale of}
\begin{equation}
\tau_\mathrm{p} \sim {P_\mathrm{o}^2 \over P_\mathrm{s}} \label{t_p}.
\end{equation}
$P_\mathrm{o}$ is the orbital period and $P_\mathrm{s}$ is the spin period of the asteroid. The trajectory of ‘Oumuamua is uncertain due to limited data and gravitational scattering. {If we substitute the orbital period $P_\mathrm{o} \approx 2.5 \times10^8$ years (the same order of magnitude as {the rotation period of the Solar system around the center of the Galaxy}) and spin period $P_\mathrm{s} \approx 8.67$ hours into Equation (\ref{t_p}), we obtain $\tau_\mathrm{p} \approx 6 \times 10^{10}$\,Gyrs. As can be seen, this precession effect is negligible in the rotation evolution. Note that stellar perturbations and the spin-orbit resonance effect are not taken into account, which may cause a slight precession, so the estimate is very rough and may be inaccurate by a few orders of magnitude. However, in view of the extremely large resulting timescale $6 \times 10^{10}$\,Gyrs, these effects are believed to be unimportant.}

The tidal effect due to a close encounter to a planetary system is believed to be unimportant since the possibility of encounters to a planetary system is extremely small due to the low stellar density for the solar neighborhood. In the case of ‘Oumuamua, the mean travel time in interstellar space before it gets close to a planetary system is estimated to $10^4$ Gyrs \citep{Y2017}.

As for the YORP effect, it is also believed to be negligible due to the very dim starlight in interstellar space \citep{HLLC2018}. Even for interstellar asteroids that entering a planetary system (e.g. ‘Oumuamua), the time they spend in the YORP-dominated zone is {hundreds to thousands} of years according to Figure \ref{comparison}.

\section{Other Effects of Collision with Interstellar Medium}
\label{B}
Interstellar gas and dust may cause erosion and accretion of the asteroid’s surface. The impacts of particles with energy higher than a few eVs can transfer energy that is higher than the lattice binding energy to the surface, therefore causing the ejection of the atoms on the surface, which is known as {``sputtering”} \citep{F1980}. This process, together with evaporation due to instantaneous high temperature \citep{S1986,BL2018} leads to the erosion on the surface. In addition, the impact on the surface by interstellar gas will cause the accretion of surface controlled by a sticking factor $f$. Therefore, such a process of accretion and erosion will contribute to the shape change of asteroids. \citet{S1986} derived the formula of the mass change caused by accretion:
\begin{equation}
dM_\mathrm{accretion} = \bar{\mu}_\mathrm{p} \bar{f} \pi R^2\bar{v} dt.
\end{equation}
Here $\bar{\mu_\mathrm{p}}$ is the mass density of the proton in the interstellar space, $\bar{f}$ is the average sticking ratio, $R$ is the equivalent radius of the projected area perpendicular to the direction of incident gas particles, $v$ is the average speed of the object with respect to the interstellar medium and $t$ is the interaction time. As for the mass change caused by erosion, \citet{VM2019} shows that the interstellar object will lose 10\,mm per million years, requiring the grain mass is about $1.5 \times 10^{-5}$\,grams. However, \citet{LBGKL2000} point out that the number density of interstellar grains in the Solar system is dominated by grains with masses between $10^{-14}$ grams and $10^{-12}$ grams. In this study I follow the formula derived by \citet{S1986}, using the parameter mean mass density  to estimate the erosion: 
\begin{equation}
dM_\mathrm{erosion} = -\bar{\mu}_\mathrm{gr}(\alpha + \beta) \pi R^2\bar{v}dt.
\end{equation}
 {\citet{LBGKL2000} estimated the total mass density of interstellar grains in the Solar system to be $6.2 \times 10^{-24}\kg\m^{-3}$ from the Ulysses and Galileo in-situ data.} $\alpha$ is the ratio of evaporated mass of the object to the mass of incident grain, $\beta$ is the ratio of the ejected mass of the object to the mass of incident grain. Here if we consider $d M = dM_\mathrm{accretion}+dM_\mathrm{accretion} = 4 \pi \rho R^2 dR$ for a rough estimation, we can obtain
\begin{equation}
{dR \over dt} \sim {{\bar{\mu}_\mathrm{p} \bar{f} \bar v - \bar{\mu}_{\rm gr} (\alpha + \beta)\bar v } \over 4 \rho}.
\end{equation}
Here I take $1\cm^{-3}$ to be the averaged number density of protons as explained in Section \ref{intro}. Based on the results from \citet{S1986}, we estimate $\bar{f} = 0.037$, $\alpha$ = 2, and $\beta$ is given by $\beta = 5.5 \times 10^{-10} \bar{v}^2$ for basalt targets. In the case of ‘Oumuamua, we take the average velocity relative to interstellar medium v as $20\km\s^{-1}$, thus $\beta$ = 2200. The equivalent radius of ‘Oumuamua is around 60m. Then we have the radius change rate is $dR/dt \approx -0.0014\m\,\rm{Gyr}^{-1} $. Thus, on the timescale considered in this paper, the shape change caused by accretion and erosion can be neglected.

On the other hand, the process of sputtering and evaporation transfers the angular momentum. \citet{S1990} has estimated the speed of ejected particles on icy surface $ 0.05\km\s^{-1}<v_{\rm ej}<0.2\km\s^{-1}$ when the incident particles have a speed of $20 \km\,\s^{-1}$. The net torque along $z$-axis is expressed as:
\begin{equation}
T_z = {-dM_\mathrm{erosion} \over dt}v_{\rm ej}RC_z = \bar{\mu}_{\rm gr} (\alpha + \beta)\bar v  v_{\rm ej} \pi R^3 C_z.
\end{equation}
Again, $C_z$ is estimated as 0.01. Substituting $v_{\rm ej} = 0.2\km\s^{-1}$ and $R = 60\m$,  we have $T_z = 3.7 \times 10^{-10}$\,N\,m. Compared with the torque produced by hydrogen atoms in Equation (\ref{T_z}), this torque is smaller by one order of magnitude and therefore gets ignored in the simulation. Note that the difference by just one order of magnitude is easily broken when considering the uncertainty of parameters. Thus the {``sputtering"} effect needs considerations when more detailed research is conducted.

\end{document}